\documentstyle[aps,prl]{revtex}
\draft

\begin{document}
\author{Daniel Boies, C. Bourbonnais and A.-M. S. Tremblay}
\title{One-particle and two-particle instability of\\ coupled Luttinger liquids
}
\date{4 July 1994}
\address{D\'epartement de Physique and Centre de Recherche en Physique
du Solide \\Universit\'e de Sherbrooke, Sherbrooke, Qu\'ebec, Canada J1K 2R1}
\maketitle

\begin{abstract}
It is shown that the Luttinger liquid is unstable to arbitrarily small
transverse hopping. It becomes either a Fermi liquid or exhibits long-range
order at zero temperature. The crossover temperatures below which either
transverse coherent band motion or long-range order start to develop can be
finite even when spin and charge velocities differ. Explicit scaling
relations for the one-particle and two-particle crossover temperatures are
derived in terms of transverse hopping amplitude, spin and charge velocities
as well as anomalous exponents. The special case of infinite-range
transverse hopping can be treated exactly and yields a Fermi liquid down to
zero temperature, unless the anomalous exponent $\theta $ is larger than
unity.
\end{abstract}

\preprint{CRPS-94-10}

The one-dimensional electron gas with short-range interactions provides the
best understood example of interacting electrons whose asymptotic low-energy
behavior is {\it not} described by a Fermi liquid (FL) fixed point. Instead,
the stable fixed point is the `Luttinger liquid' (LL). From the time of
their discovery, high-temperature superconductors lead to question the
stability of the FL fixed point in {\it two} dimensions. In particular,
Anderson \cite{Anders} has put forward the idea that a Luttinger liquid
could survive in two dimensions. It became clear that Luttinger liquids
weakly coupled by interchain single particle hopping $t_{\bot }$ would be a
natural model to give a sound basis to this conjecture. In fact, the
question of the stability of the Luttinger liquid in the presence of $%
t_{\bot }$ has been addressed long ago in the context of
quasi-one-dimensional conductors \cite{Bourbon1,Brazo,Bourbon2}. It was then
shown that $t_{\bot }$ as well as pair-hopping correlations destabilize the
Luttinger liquid fixed point. A series of recent works on two coupled chains
\cite{Twochain} and on the many-chain problem \cite{Castel92} confirm this
view but, nevertheless, the issue remains controversial \cite{Anders3}. In
particular, going beyond the two chain case is a necessary requirement for
phase transitions or long-range quasi-particle coherence to occur. The very
few attemps to do so essentially deal with the situation where the spin and
charge velocities are equal \cite{Brazo,Bourbon2}.

In this Letter we address the problem of interchain coherence for an
infinite number of coupled chains starting from a new functional integral
formulation where $t_{\bot }$ is the perturbation and the unperturbed system
is the LL with anomalous exponents {\it and} differing spin and charge
velocities. As shown at the end of this letter, our approach becomes exact
in the limit of infinite-range transverse hopping. We investigate both
single-particle spectral weight and induced two-particle correlations. For a
given bare $t_{\bot }$ and temperatures much lower than the Fermi energy $%
T\ll E_F$, the typical phase diagram found here is sketched in Fig.1 as a
function of temperature and of the anomalous exponent $\theta $ which is a
measure of the interaction strength. As temperature is lowered, two types of
crossovers can occur. For weak enough interaction, transverse one-particle
coherent motion starts to develop and the system crosses over from LL to FL
behavior at the deconfinement temperature $T_{x^1}$. On the other hand, for
strong interaction, virtual pair hopping becomes the dominating process
which eventually leads to long-range ordering below the two-particle
dimensional crossover temperature $T_{x^2}$, even if there is confinement at
the one-particle level ($T_{x^2}\negthinspace >\negthinspace T_{x^1}$). At
temperatures such that $0<$$T\negthinspace <T_{x^1}\left( T_{x^2}\right) $,
a true phase transition can occur in more than two dimensions.

Let us start with the full partition function for a set of $N_{\bot }$ fully
interacting chains written in the interaction representation
\begin{equation}
{Z={\rm Tr}\{e^{-\beta \left( \sum_i{{\cal H}_i^{1D}}+\sum_{ij}{{{\cal H}%
_{\bot ij}}}\right) }\}=Z_{1D}\langle {T_\tau e^{-\int_0^\beta d\tau {%
\sum_{ij}}{\cal H}_{\bot ij}\left( \tau \right) }}\rangle _{1D}}
\end{equation}
where indices $i,j$ run over all $N_{\bot }$ chains, ${\cal H}_i^{1D}$ is
the purely $1D$ Hamiltonian describing the interacting electrons along chain
$i$ while the interchain hopping part ${{{{\cal H}_{\bot ij}}}}$ stands as
the perturbation. The above thermodynamic average $\langle {\ldots }\rangle
_{1D}$ and~partition function~$Z_{1D}$ only involve the pure $1D$
Hamiltonian. The hopping Hamiltonian is given by ${{\cal H}_{\bot ij}=-\int
dx\sum_{p,\sigma }t_{\bot ij}{a_{p,i}^{\sigma \dagger }}(x)a_{p,i}^\sigma (x)%
}$ where $x$ is the continuous coordinate along the chains while $p=\pm $
denote right and left-going electrons respectively. By analogy with the
problem of propagation of correlations in critical phenomena, the
propagation of one-particle transverse coherence is studied through an
effective field theory which is generated by a Hubbard-Stratonovich
transformation for Grassmann variables. This allows the partition function $%
Z $ to be expressed as a functional integral over a Grassmann $\psi ^{\left(
*\right) }$ field\widetext
\begin{equation}
\label{un}Z=Z_{1D}\negthinspace \int \negthinspace \negthinspace
\negthinspace \int {\cal D}\psi ^{*}{\cal D}\psi \,e^{\;-\int d(1)%
\negthinspace \sum\limits_{ij}\psi _i^{*}(1)t_{\bot ij}^{-1}\psi _j(1)}{%
\langle \;e^{\;\int d(1)\negthinspace \sum\limits_i\left( a_i^{\dagger
}(1)\psi _i(1)+\psi _i^{*}(1)a_i(1)\right) }\;\rangle _{1D}}
\end{equation}
with the notations $\int d(n)\psi _i(n)\equiv \int
dz_n\sum\nolimits_{p_n,\sigma _n}\psi _{p_n,i}^{\sigma _n}(z_n)$ and $\int
dz_n\equiv \int dx_n\int_0^\beta d\tau _n$. The thermodynamic average in (%
\ref{un}) is readily recognized as the generating function for the exact $1D$
connected Green's functions $G_c^{(n)}$. Changing variables from $\psi _i$
to $\sum_jt_{\bot ij}^{1/2}\psi _j$ in the functional integral (\ref{un})
allows to ultimately write for the effective field theory
\begin{equation}
\label{deux}{Z}=Z_{1D}\int \negthinspace \negthinspace \negthinspace \int
{\cal D}\psi ^{*}{\cal D}\psi \,e^{\,{\cal F}[\psi ^{*},\psi ]}
\end{equation}
where the Grassmanian Landau-Ginzburg-Wilson functional ${\cal F}={\cal F}%
_0+\sum\nolimits_{n\ge 2}{\cal F}^{(n)}$ involves a quadratic part and a sum
of effective interactions to all orders in the $\psi $ field. To write a
specific form for ${\cal F}$, let us consider the case where chains are
lined up in a plane and let us Fourier transform in the direction transverse
to the chains. Then, the Gaussian part describing the free propagation of
the $\psi $ field, takes the form \widetext
\begin{equation}
\label{trois}{\cal F}_0=-\int \negthinspace d(1)\negthinspace \int
\negthinspace d(2)\negthinspace \sum\limits_{k_{\bot }}\psi _{k_{\bot
}}^{*}(1)\left( \openone -t_{\bot }(k_{\bot })G^{(1)}(1-2)\right) \psi
_{k_{\bot }}(2),
\end{equation}
where $G^{(1)}$ is the exact one-dimensional propagator. Also, the
interacting part is found to be
\begin{equation}
\label{quatre}{\cal F}^{(n)}={\frac 1{n!}}{\frac 1{N_{\bot }^{n-1}}}%
\sum\limits_{k_{\bot 1}\ldots k_{\bot 2n}}\negthinspace \int \negthinspace
d(1)\ldots \int \negthinspace d(2n)\psi _{k_{\bot 1}}^{*}(1)\ldots \psi
_{k_{\bot 2n}}(2n)\gamma _{\bot }\left( k_{\bot 1}\ldots k_{\bot 2n}\right)
G_c^{(n)}(1\ldots 2n)
\end{equation}
where
\begin{equation}
\label{gamma}\gamma _{\bot }\left( k_{\bot 1}\ldots k_{\bot 2n}\right)
=\left\{ \prod_\alpha ^{2n}t_{\bot }\left( k_{\bot \alpha }\right) \right\}
^{1/2}\delta _{\Sigma _\alpha k_{\bot \alpha },0}\ \;
\end{equation}
in which $t_{\perp }(k_{\perp })$ is the Fourier transform of $t_{\perp ij}$%
. In the non-interacting limit, $G_c^{(n\geq 2)}\rightarrow 0$ in which case
(\ref{deux}) correctly reduces to the $2D$ free-electron partition function.

{\it Quasi-particle pole at }${\it T\negthinspace =\negthinspace 0}.$ An
instability of the LL at zero temperature and thus the possibility of a FL
fixed point is already present in the free theory of the $\psi $ field
described by ${\cal F}_0$. At this level of approximation, the $2D$
one-particle Green's function in Fourier-Matsubara space, say for
right-going electrons, is given by
\begin{equation}
\label{six}{\cal G}^{2D}({\bf k},\omega _n)={\frac{{G^{(1)}(k,\omega _n)}}{%
1-t_{\bot }(k_{\bot })G^{(1)}(k,\omega _n)}},
\end{equation}
where $\omega _n\negthinspace =\negthinspace (2n+1)\pi T$, while ${\bf k}%
=(k,k_{\bot })$ with $k$ measured with respect to the right $1D$ Fermi point
$k_F^0$. In order to prove the existence of a quasi-particle pole at $T%
\negthinspace =\negthinspace 0$, we use the known general form for the
asymptotic $1D$ propagator $G^{(1)}$ describing the Luttinger liquid in
space and imaginary time \cite{Korep90},
\begin{equation}
\label{cinq}G^{(1)}(x,\tau )={\frac{e^{ik_F^0x}}{{2\pi i\Lambda ^\theta }}}%
\prod_{\nu =\rho ,\sigma }S(z_\nu )^{-1/2-\theta _\nu /2}S(z_\nu
^{*})^{-\theta _\nu /2}
\end{equation}
where $S(z_\nu )=\xi _\nu \sinh \left( {z_u/\xi _\nu }\right) $ and $z_\nu
\negthinspace =\negthinspace x\negthinspace +\negthinspace iv_\nu \tau $.
Here, $v_\rho $ and $v_\sigma $ are the velocities of spin and charge
excitations respectively while $\xi _{\rho ,\sigma }=v_{\rho ,\sigma }/\pi T$
are the corresponding thermal coherence lengths. The exponents $\theta _\rho
>0$ and $\theta _\sigma >0$ are the spin and charge contributions to the
anomalous dimension $\theta \negthinspace =\negthinspace \theta _\rho
\negthinspace +\negthinspace \theta _\sigma $ of the $1D$ Green's function
and $\Lambda $ is an ultraviolet cutoff. Now, one of the central issues is
to shed light on the influence of differing spin and charge velocities \cite
{Anders} on the stability of the LL. For this sake, let us consider (\ref
{cinq}) in a special case where things get simpler, that is, when $\theta
_\rho \negthinspace =\negthinspace \theta _\sigma \negthinspace
=\negthinspace 0$ but still $v_\rho \negthinspace \neq \negthinspace
v_\sigma $. In this case, the $T\negthinspace =\negthinspace 0$ retarded $1D$
propagator $G^{(1)}(k,\omega )$ has two square root singularities. At the
Gaussian level, the corresponding spectral weight $A^{2D}=-\pi ^{-1}{\rm Im}%
{\cal G}^{2D}$ obtained from the imaginary part of (\ref{six}) is given by
\begin{equation}
\label{neuf}A^{2D}({\bf k},\omega )={\frac{A^{1D}(k,\omega )}{{1+\left( \pi
t_{\bot }(k_{\bot })A^{1D}(k,\omega )\right) ^2}}}+Z(t_{\bot })\delta
(\omega -\epsilon ({\bf k))}
\end{equation}
where $A^{1D}=-\pi ^{-1}$Im$G^{\left( 1\right) }$ is the exact
one-dimensional spectral weight, $Z\left( t_{\bot }\right) $ is the
quasi-particle residue and $\epsilon \left( {\bf k}\right) $ the pole of (%
\ref{six}). For the case at hand, the undamped quasi-particle spectrum given
by the pole of (\ref{six}) is $\epsilon \left( {\bf k}\right) =k\overline{v}$
$-{\rm sign}\left\{ t_{\bot }\left( k_{\bot }\right) \right\} \sqrt{\left(
k\Delta v\right) ^2+t_{\bot }^2\left( k_{\bot }\right) }$ where $\overline{v}%
\negthinspace =\negthinspace \left( v_\rho +v_\sigma \right) /2$ and $\Delta
v\negthinspace =\negthinspace \left( v_\rho -v_\sigma \right) /2$, while the
quasi-particle residue takes the form

\begin{equation}
\label{dix}Z(t_{\bot })={\frac{{\left| \,t_{\bot }(k_{\bot })\right| }}{{\ }
\sqrt{{(k\Delta v)^2+t_{\bot }^2(k_{\bot })}}}}\text{.}
\end{equation}
The residue is readily seen to satisfy $Z\negthinspace \rightarrow
\negthinspace 0$ when $t_{\bot }\negthinspace \rightarrow \negthinspace 0$
while $Z\negthinspace \rightarrow \negthinspace 1$ when $\Delta v%
\negthinspace \rightarrow \negthinspace 0$ as one expects for free
electrons. Note that the infinite lifetime of the quasi-particles in (\ref
{neuf}) should become finite when one goes beyond the Gaussian approximation
thus allowing the $\psi ^{\prime }s$ to interact, except at the Fermi
surface where the lifetime should remain infinite because of the usual phase
space arguments. One can also check that for wave-vectors close to the new
Fermi surface, given by $(v_\rho v_\sigma )^{1/2}k=t_{\bot }(k_{\bot })$,
the single-particle spectral weight has the following frequency dependence.
For nearest-neighbor hopping and $k_{\bot }\negthinspace <\negthinspace \pi
/2$, a quasi-particle peak is first encountered as the frequency is
increased, followed by an incoherent background which is a smoothed version
of the original LL. In other words, remnants of spin-charge separation are
left at high energies.

{\it One-particle dimensional crossover. }In the more general case $\theta
\negthinspace \neq \negthinspace 0$, $Z\left( t_{\bot }\right) $ cannot be
found analyticaly although we can prove from known $1D$ spectral weights
\cite{Voit} that a pole appears in regions where the $1D$ spectral weight is
zero, leading to results qualitatively similar to those just discussed.
Nevertheless, we can determine the temperature scale $T_{x^1}$ at which the
FL pole in ${\cal G}^{2D}$ becomes perceptible and transverse single-fermion
coherence starts to develop. Using the natural change of variables $\tau
^{\prime }=\pi T\tau $ and $x=(\xi _\rho \xi _\sigma )^{1/2}x^{\prime }$ to
evaluate the Fourier transform of the $1D$ Green's function (\ref{cinq}) at
the Fermi level $G^{(1)}(0,\pi T)$ , and substituting in the $2D$ Gaussian
propagator (\ref{six}) one readily finds
\begin{equation}
\label{Tx1}T_{x^1}\sim E_F\left( {\frac{t_{\bot }}{E_F}}\right) ^{1/\left(
1-\theta \right) }\left( {\frac{v_F}{v_\rho }}\right) ^{\theta _\rho /\left(
1-\theta \right) }\left( {\frac{v_F}{v_\sigma }}\right) ^{\theta _\sigma
/\left( 1-\theta \right) }F_1\left( \left\{ {v_\sigma /v_\rho }\right\}
^{1/2}\right)
\end{equation}
where $F_1\left( x\right) $ is a temperature independent function that
satisfies $F_1\left( x\right) =F_1\left( 1/x\right) $and which also depends
on $\theta _{\rho ,\sigma }$. As seen from (\ref{Tx1}), the effect of the
difference in the spin and charge velocities is to decrease the
deconfinement temperature but not to make it vanish. The vanishing of $%
T_{x^1}$ is expected for sufficiently strong coupling since spin and charge
degrees of freedom must recombine for an electron to tunnel on a
neighbouring chain. Indeed, it can be shown that $F_1\left( \left\{ v_\sigma
/v_\rho \right\} ^{1/2}\right) \stackrel{v_\sigma /v_\rho \rightarrow 0}{%
\sim }\left\{ v_\sigma /v_\rho \right\} ^{1/2}$ and correspondingly, $%
T_{x^1}\rightarrow 0$ when $\theta _\sigma <\frac 13(1-\theta _\rho )$.
Following the standard notation for the crossover temperature $T_{x^1}\sim
t_{\bot }^{1/\phi _{x^1}}$, the single particle crossover exponent is $\phi
_{x^1}=1\negthinspace -\negthinspace \theta $ as already found elsewhere
\cite{Bourbon1,Bourbon2,Castel92}. Consequently, as long as $\theta
\negthinspace
<\negthinspace 1$, or equivalently if $G^{(1)}(x)$ decays more slowly than $%
x^{-2}$, the coupling $t_{\bot }$ is relevant and $T_{x^1}$ is finite,
although smaller than the non-interacting value $T_{x^1}\sim t_{\bot }/\pi $%
. The condition $\theta \negthinspace <\negthinspace 1$ is satisfied for the
Hubbard model with a non half-filled band where one has the exact result $%
\theta \negthinspace \le \negthinspace {1/8}$ \cite{Schulz}. For more
specialized 1D models (forward scattering only, half-filling, etc) one can
have $\theta \negthinspace =\negthinspace 1$ and $\theta \negthinspace
>\negthinspace 1$ where transverse hopping becomes marginal and irrelevant
respectively. In these cases, transverse band motion does not develop and
the electrons remain spatially confined along the chains at all temperatures.

{\it Two-particle dimensional crossover.} We now proceed beyond the Gaussian
level by taking into account the ${\cal O}(t_{\bot }^2)$ quartic term in the
functional (\ref{quatre}) which describes correlated transverse pair
tunneling. We argue that the system will undergo a two-particle dimensional
crossover towards CDW, SDW ordered states if the interaction is repulsive
and singlet or triplet superconducting states if it is attractive. Focussing
on the $2k_F^0$ particle-hole channel, we rewrite the partition function at
the quartic level
$$
{Z}=Z_{1D}\langle {e^{\sum\limits_{\mu ,q_{\bot }}\int \{dz\}O_{\mu ,q_{\bot
}}^{*}(z_3,z_1)\gamma _{\bot }(q_{\bot })R_\mu (\{z\})O_{\mu ,q_{\bot
}}(z_2,z_4)}}\rangle _{\psi ^{*}\psi }
$$
with the obvious notation $\{z\}\negthinspace =\negthinspace
\{z_1,z_2,z_3,z_4\}$ and where $\langle \ldots \rangle _{\psi ^{*}\psi }$ is
the average with respect to ${\cal F}_0$. The composite fields ${\it O}_{\mu
,q_{\bot }}(z_3,z_1)=N_{\bot }^{-{1/2}}\sum_{k_{\bot }}\sum_{\alpha \beta
}\psi _{-,k_{\bot }}^{\alpha *}(z_3)\sigma _\mu ^{\alpha \beta }\psi
_{+,k_{\bot }+q_{\bot }}^\beta (z_1)$ describe CDW ($\mu \negthinspace
=\negthinspace 0$) and SDW ($\mu \negthinspace =\negthinspace 1,2,3$)
correlations. Considering the case of nearest-neighbor hopping, we
approximate the transverse pair tunneling amplitude as $\gamma _{\bot
}(q_{\bot })\approx (2t_{\bot })^2\cos (q_{\bot })$ where $q_{\bot }$ is the
transverse momentum of the particle-hole pair, by setting the incoming
momenta to $0$ or $\pi $ since this leads to the highest value for $T_{x^2}$%
. In the above, $\sigma _0^{\alpha \beta }=\delta ^{\alpha \beta }$, $\sigma
_{\mu =1,2,3}^{\alpha \beta }$ are the Pauli matrices and the $q_{\bot }$
independent function $R_\mu (\{z\})$ is the $1D$ connected correlator for
charge or spin fluctuations.

To examine the possibility of phase transition, we perform a
Hubbard-Stratonovitch transformation on the ${\it O}_\mu $ fields. Let $%
\zeta _{\mu ,q_{\bot }}^{*}(z_3,z_1)$ be the complex field conjugate to $%
{\it O}_{\mu ,q_{\bot }}(z_3,z_1)$. The partition function $Z$ then takes
the form%
\begin{eqnarray}
Z=Z_{1D}\int\!\!\!\!\int{\cal D}\zeta^*{\cal D}\zeta\,
e^{-\;\int\! \{dz\}\! \sum\limits_{\mu,q_{\bot}}
\zeta_{\mu,q_{\bot}}^{*}(z_3, z_1)
\bigl(\openone-\gamma_{\bot}(q_{\bot}){\mit R}_{\mu}(\{z\})\bigr)
\zeta_{\mu,q_{\bot}}(z_2, z_4)+{\cal O}(\zeta^4)}.
\end{eqnarray}
Softening of the $\zeta $ field first occurs for a value of $q_{\bot }%
\negthinspace =\negthinspace \pi $ corresponding to the ususal staggered
order. The temperature $T_{x^2}$ at which the $\zeta $ field softens must be
greater than $T_{x^1}$ to retain its meaning as a two-particle crossover
temperature. For definiteness, let us now consider the gapless case. As
shown below, $T_{x^2}$ scales differently in the weak- and strong-coupling
limits. In the strong-coupling limit, the correlator $R_\mu
(z_1-z_3,z_2-z_4,z_1-z_2)$ decays faster than the square of the
electron-hole separations $|z_1-z_3|$ and $|z_2-z_4|$ so that the scaling is
determined by $|z_1-z_2|$ only, leading to the asymptotic form
$$
R_\mu (z=z_1-z_2)\sim -\Lambda ^{\gamma _\mu }\cos 2k_F^0x\prod_{\nu =\rho
,\sigma }\left[ f_\nu (z_\nu ,z_\nu ^{*})S(z_\nu )S(z_\nu ^{*})\right]
^{-\gamma _{\mu ,\nu }/2}
$$
with $\gamma _\mu =2-\gamma _{\mu ,\rho }-\gamma _{\mu ,\sigma }$ and $f_\nu
(z_\nu ,z_\nu ^{*})=2\mid z_\nu \mid ^2[z_\nu ^2+z_\nu ^{*2}]^{-1}$. The
strong-coupling regime is defined by $2-2\theta -\gamma _\mu <0$ \cite
{Brazo,Bourbon2} where $\gamma _\mu $ is the critical exponent which governs
the temperature behavior of the $1D$ susceptibility {$\chi _\mu ^{1D}(T)\sim
T^{-\gamma _\mu }$}. In this regime,
\begin{equation}
\label{Tx2a}T_{x^2}^\mu \sim E_F\left( {\frac{t_{\bot }^2}{E_F^2}}\right)
^{1/\gamma _\mu }\left( {\frac{v_F}{v_\rho }}\right) ^{\left( 2\gamma _{\mu
,\rho }-1\right) /2\gamma _\mu }\left( {\frac{v_F}{v_\sigma }}\right)
^{\left( 2\gamma _{\mu ,\sigma }-1\right) /2\gamma _\mu }F_2^\mu \left(
\left( v_\sigma /v_\rho \right) ^{1/2}\right) \;
\end{equation}
where the dimensionless function $F_2^\mu (x)$ vanishes at $x=0$ and $%
E_F\sim $ $\Lambda v_F$ .

In the weak-coupling regime, $2-2\theta -\gamma _\mu >0$, the correlator
essentially scales as the square of the one-particle propagator (\ref{cinq})
so that
\begin{equation}
\label{Tx2b}T_{x^2}^\mu \sim T_{x^1}F_3^\mu \left( \left( v_\sigma /v_\rho
\right) ^{1/2},\left\{ g\right\} \right) .
\end{equation}
where $\left\{ g\right\} $ stands for the set of dimensionless
electron-electron interaction vertices and $F_3^\mu (x,y)$ vanishes when
either of its arguments vanishes. It is only for sufficiently weak coupling
that $T_{x^2}<T_{x^1}$ and that deconfinement takes place. In both regimes,
the effect of $v_\rho \negthinspace \neq \negthinspace v_\sigma $ is to
decrease $T_{x^2}$ . Finally, (\ref{Tx2a}) and (\ref{Tx2b}) reduce to the
known results \cite{Bourbon2} when $v_\rho \negthinspace =\negthinspace %
v_\sigma $.

{\it Infinite range transverse hopping: an exact result.} The case where the
transverse hopping amplitude has an infinite range can be solved exactly
\cite{These}. For the energy to be independent of the number of
perpendicular chains $N_{\bot }$ in the thermodynamic limit $N_{\bot }%
\negthinspace \rightarrow \negthinspace \infty $, the transverse hopping
matrix must be scaled as $t_{\bot ij}={t_{\bot }/N_{\bot }}$. This implies
that in Fourier space, $t_{\bot }(k_{\bot })=t_{\bot }\delta _{k_{\bot },0}$
so the $n$-body interaction is such that ${\cal F}^{(n\geq 2)}\propto
t_{\bot }\left( t_{\bot }/N_{\bot }\right) ^{n-1}$. Consequently, in the
limit $N_{\bot }\negthinspace \rightarrow \negthinspace \infty $, all
effective interactions vanish as $t_{\bot }^n/N_{\bot }^{n-1}\negthinspace %
\rightarrow \negthinspace 0$. Hence, the Gaussian propagator (\ref{six})
becomes exact. There can be {\it no} interchain pair tunneling and
consequent long-range order. Either $\theta \ge 1$ and the electrons remain
spatially confined along the chains at all temperatures, or $\theta <1$ and
the system remains a Fermi liquid at ${\bf k}=(k,0)$ for all temperatures
below $T_{x^1}$ while for other values of ${\bf k}$, $t_{\perp }(k_{\perp })$
has no effect and the electrons remain confined.

We wish to thank M. Fabrizio and J. Voit for helpful discussions. We
acknowledge financial support from the Natural Sciences and Engineering
Research Council of Canada (NSERC), the Fonds pour la formation de
Chercheurs et l'aide \`a la Recherche from the Government of Qu\'ebec
(FCAR), the Canadian Institute of Advanced Research (CIAR) and (A.M.S.T.)
the Killam foundation.

\begin{figure}
\caption{One- and two-particle crossover
temperatures shown qualitatively as a
function of $\theta$ in the gapless case. The dotted (dash-dotted)
line indicates the crossover LL$\rightarrow$FL (FL$\rightarrow$LRO)
for the exact infinite-range transverse hopping model (below $T_{x^1}$).
Point A gives the non-interacting value of $T_{x^1}$ ($\sim t_{\bot}/\pi$)
when $v_{\rho}=v_{\sigma}=v_F$.}
\label{autonum}

\end{figure}

\end{document}